\documentclass[aps,pre,twocolumn,showpacs,preprintnumbers,amsmath,amssymb]{revtex4-1}

\usepackage{graphicx}
\usepackage{dcolumn}
\usepackage{bm}
\usepackage[latin1]{inputenc}
\usepackage{amsfonts}
\usepackage{amsmath}
\usepackage{amssymb}
\usepackage{amsthm}
\usepackage{bbold}
\usepackage{color}
\usepackage[american]{babel}
\usepackage[T1]{fontenc}
\usepackage{longtable}
\usepackage{xspace}
\usepackage{bbm}
\usepackage[all]{xy}

\newcolumntype{C}[1]{>{\centering\arraybackslash}p{#1}}

\begin{document}


\title{Nuclear quantum effects on the vibrational dynamics of  the water-air interface}

\author{Deepak Ojha}
\author{Andr\'es Henao}
\author{Frederik Zysk}
\affiliation{%
Dynamics of Condensed Matter and Center for Sustainable Systems Design, Department of Chemistry, University of Paderborn, Warburger Str. 100, D-33098 Paderborn, Germany
}
\author{Thomas D. K\"uhne}
\email{tkuehne@cp2k.org}
\affiliation{%
Dynamics of Condensed Matter and Center for Sustainable Systems Design, Department of Chemistry, University of Paderborn, Warburger Str. 100, D-33098 Paderborn, Germany
}
\affiliation{Paderborn Center for Parallel Computing and Institute for Lightweight Design, University of Paderborn, Warburger Str. 100, D-33098 Paderborn, Germany}

\date{\today}

\begin{abstract}

We have applied path integral molecular dynamics simulations to investigate nuclear quantum effects on the vibrational dynamics of water molecules at the water-air interface.
The instantaneous fluctuations in the frequencies of the O-H stretch modes are calculated using the wavelet method of time series analysis, while the time scales of vibrational spectral diffusion are determined from frequency-time correlation functions and joint probability distributions, as well as the hydrogen bond number correlation functions. We find that the inclusion of nuclear quantum effects leads not only to a redshift in the vibrational frequency distribution by about 120~cm$^{-1}$ for both the bulk and interfacial water molecules, but also to an acceleration of the vibrational dynamics on the water-air interface by as much as 60$\%$. In addition, a blueshift of about 30 ~cm$^{-1}$ is seen in the vibrational frequency distribution of interfacial water molecules compared to that of the bulk. Furthermore, the dynamics of water molecules beyond the topmost two molecular layers was found to be essentially similar to that of bulk water.
\end{abstract}

\maketitle

\section{Introduction}
Understanding the behavior of water at interfaces is a primary interest of the biological, chemical and earth sciences \cite{ball,ball2008,kauzmann}. An asymmetrical solvent environment exists at the interface, which leads to variation in the dynamics, structure, reactivity and selective
adsorption at the water-air interface \cite{tdk_1,tdk_3,tdk_4,skinner,richmond,richmond2,richmond3}. This asymmetrical local solvent structure at the interface is characterized by the hydrogen bonding of the interfacial water molecules. The temporal and spatial reorganization of the local hydrogen bond network is intimately connected with the instantaneous fluctuations in vibrational frequencies of the OH/OD modes in liquid water \cite{skinner2,tdk_sr}.  Recent experimental techniques like
vibrational sum-frequency generation (vSFG) spectroscopy measure the spectral contributions from the non-centrosymmetric region of a solution and thus
has enabled a direct surface-specific probe of the molecular structure and dynamics of interfacial water molecules \cite{shen1, shen2,shen3,mbonn,allen,tahara}.
However, the interpretation of the observed vSFG spectrum is complicated, as there is no direct way for uniquely decomposing the observed peaks.
In this context, several molecular dynamics (MD) studies have been performed to interpret the relationship between the vibrational spectrum and the 
vibrational dynamics of interfacial water molecules \cite{paesani1,richmond4,richmond5,skinner3, skinner4,skinner5,skinner6,skinner7,morita1, morita2,yuki,tdk_cc,tdk_sr2,tdk_wp}. Previous simulation studies have shown that the hydrogen-bonded network becomes less 
structured at the interface, with an enhanced population of single donor (SD) configurations, which is basically attributed to the sharp peak around
3700 cm$^{-1}$ \cite{tdk_1,kessler}. Hynes and coworkers employed a mixed quantum/classical approach for describing vibrations and reported the first simulated vSFG
spectrum of an water-air interface \cite{morita1,morita2}. Along similar lines, Skinner and coworkers used an empirical map-based quantum/classical approach to determine the
vSFG spectrum of the water-air interface \cite{skinner3,skinner4,skinner5,skinner6,skinner7}. The surface-specific velocity-velocity correlation function approach is also extensively used to obtain the vSFG spectrum of the aqueous solutions \cite{yuki,naveen}.
Quite recently, fully DFT-based ab-initio MD studies have also reported the layer-wise signal contribution
to vSFG for the water-air interface \cite{sprik,simone}.
However, because water consists of light hydrogen atoms, the inclusion of zero-point energy and quantum tunneling
effects is important for understanding the microscopic behavior of aqueous solutions and interfaces \cite{ceriotti}.
\noindent
 \ \\

The path-integral molecular dynamics (PIMD) scheme is known to incorporate nuclear quantum effects (NQEs) within atomistic simulations \cite{hibbs,PIMD}. Earlier studies have shown that the inclusion of NQEs leads to a less structured first coordination shell accompanied by an enhanced order in the second coordination shell of liquid water \cite{ceriotti}.  Similarly, the diffusion
coefficients were reported to show an increase of 50 $\%$ owing to NQEs \cite{paesani}. However, recent PIMD-based studies have shown that a competition between the inter- and intramolecular
NQEs exists, so the overall effects can be significantly less \cite{haberson_3,tdk_5}. Interestingly, the contribution of NQEs also varies, depending on the dynamical property being investigated. While the 
reorientational relaxation times see an increment of 15 $\%$ \cite{laage}, the vibrational frequency correlation decay is accelerated by nearly 30 $\%$ \cite{tdk_6}. In the present work, we have investigated the role of NQEs on the vibrational dynamics of water molecules at the water-air interface. We have also determined the extent to which the effect of interfacial effects persists by
determining the layer-specific vibrational spectral diffusion. The timescales of decay of the frequency-time correlation function (FTCF) are correlated with the hydrogen bond reorganization using the temporal decay of  $R_{O \cdots O}$ fluctuation decay and hydrogen bond number correlation functions. Moreover, the time-dependent frequency joint probability distributions,
frequency structure correlation and vSFG spectrum are also computed.  

\section{Computational Details}
In the present study, we have conducted PIMD simulations in the canonical NVT ensemble consisting of 216 water molecules using the flexible q-TIP4P/F water model developed by Habershon and coworkers \cite{haberson_3}. Earlier studies have shown the flexible q-TIP4P-F model to be well suited to determine the impact of NQEs on the structural, thermodynamical, and dynamical, as well as spectroscopic properties of liquid water \cite{haberson_3,tdk_6,tdk_5}.
The water-air interface was generated by creating a rectangular simulation box of 18.64~\AA \ in x- and y-direction and 93.22~\AA \ in z-direction. The system was replicated periodically along all three dimensions using the minimum image convention. Short-range interactions were truncated at 9~\AA \ and the Ewald summation mehtod was employed to determine the long-range electrostatic interactions. The ring-polymer contraction scheme with a cutoff value of $\sigma=5$~\AA \, was used to reduce the computationally expensive part of electrostatic forces calculation to a single Ewald sum \cite{haberson_3}. While a $p=32$ ring polymer bead was employed, the computationally expensive electrostatic calculations were contracted to the centroid only. In contrast to the original PIMD scheme, the partially adiabatic centroid MD (PACMD) technique supports the evaluation of dynamical quantities within the PIMD framework \cite{hone}. The effective mass of the ring-polymer beads is adjusted by modifying the elements of the Parrinello-Rahman mass matrix so as to recover the correct dynamics of the centroids and allow for an integration time-step close to the ionic resonance limit. The temporal evolution of the ring-polymer was performed analytically in the normal mode representation by a multiple time-step algorithm using a discretized time-step of 0.5~fs for the intermolecular and 0.1~fs for the intramolecular interactions \cite{MTS}. The systems were initially equilibrated for  20~ps, followed by production runs of 100~ps each. For comparison, an additional simulation with classical nuclei, i.e. $p=1$ was performed.

\section{Results and Discussion}

\subsection{Structure and reorientational dynamics}
\begin{figure}
\begin{center}
\includegraphics[width=0.4875\textwidth]{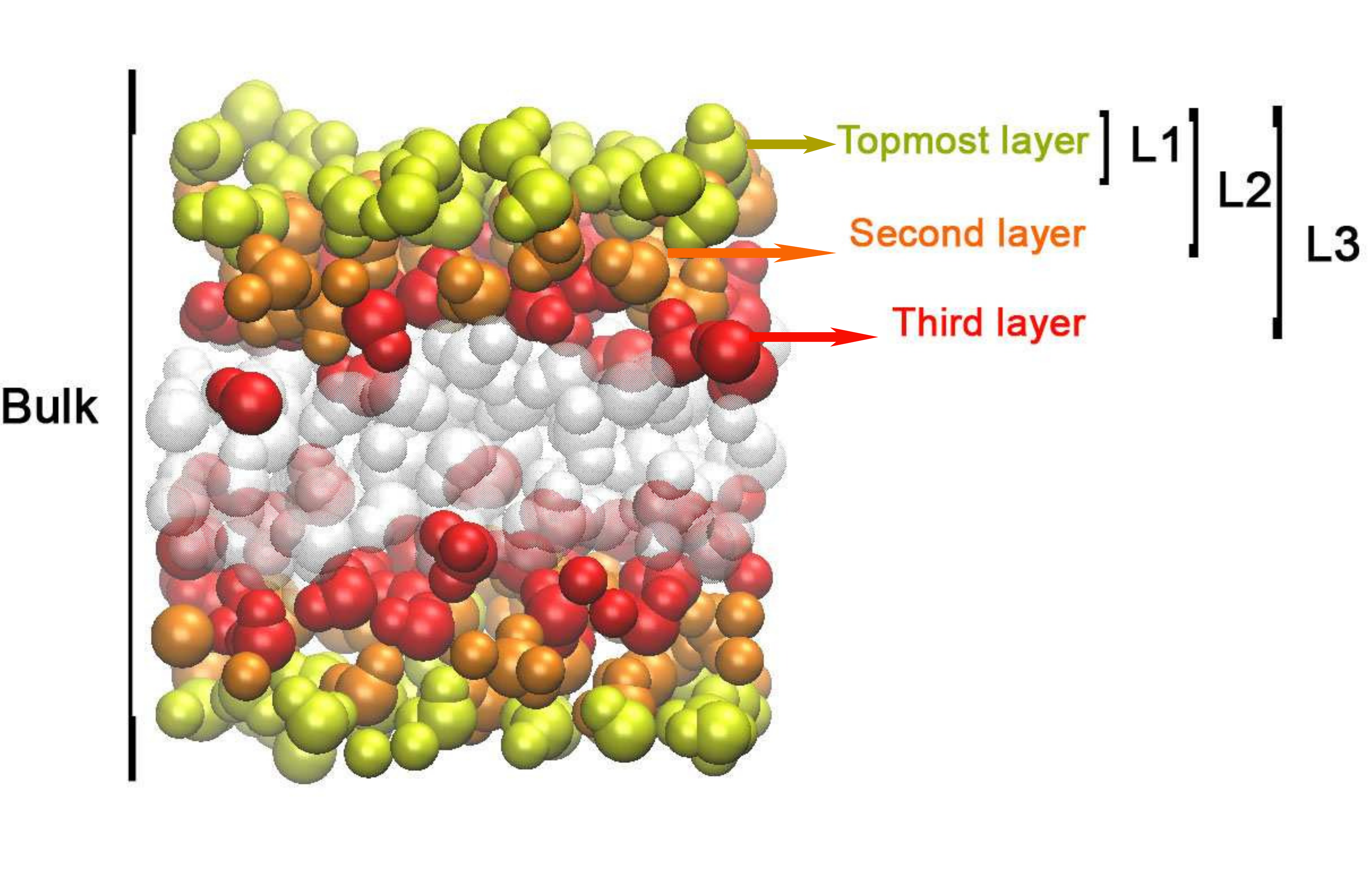}
\end{center}
\caption{\label{layers}Definition of the different layers using the ITIM  method \cite{itim,gitim}. The surface or topmost layer is first identified. Then, the procedure is repeated for the second and third layers. Finally, the inner molecules (transparent) are shown. The definition used in this work uses the topmost layer as L1, surface and second layers as L2, first three layers as L3 and, finally, the whole system is considered as bulk water. This figure was produced using VMD \cite{vmd}.
} \end{figure}

\begin{figure}
\begin{center}
\includegraphics[width=0.4875\textwidth]{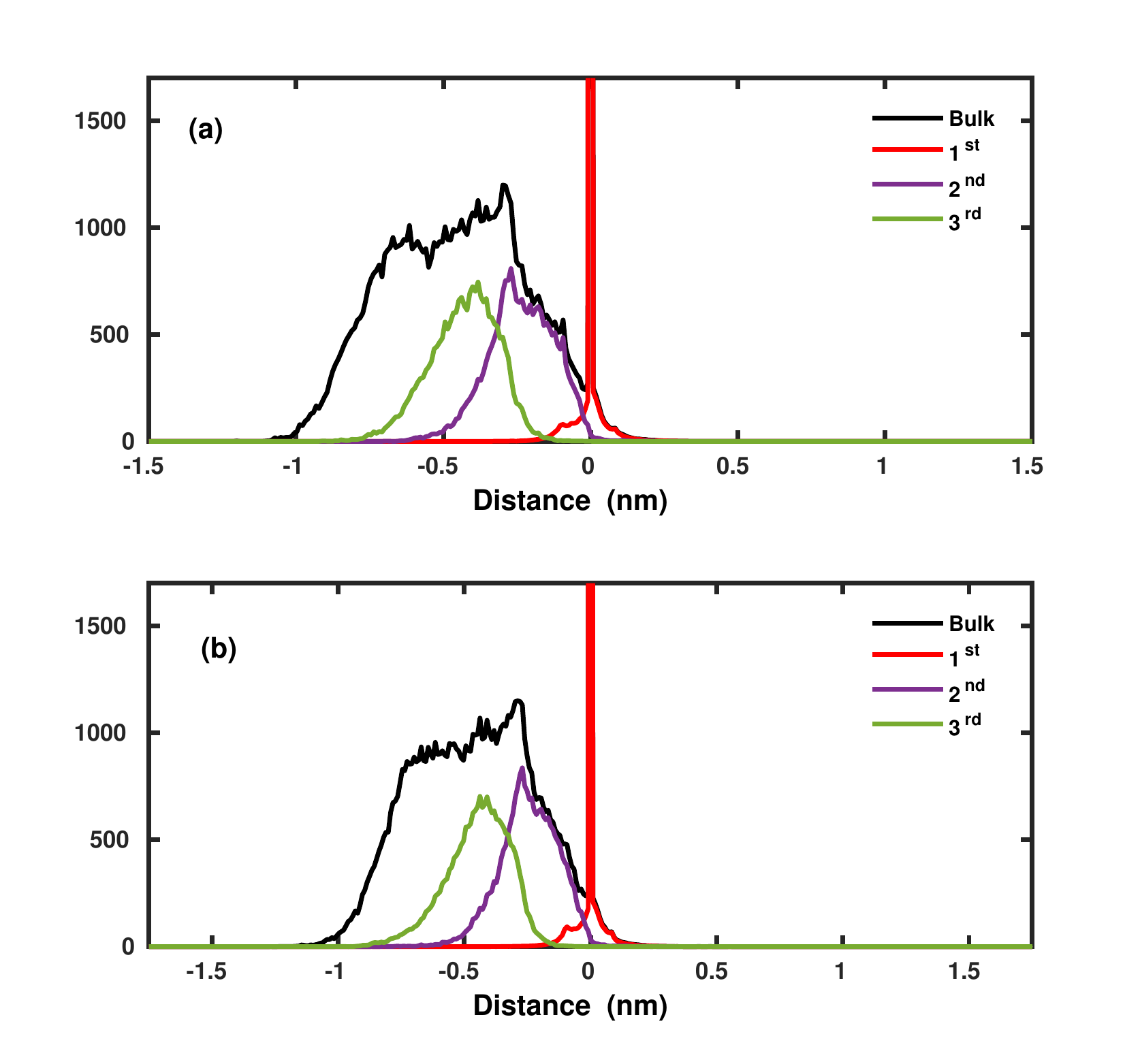}
\end{center}
\caption{\label{Fig2}
Intrinsic density profile for the molecular layers on the surface as obtained from our classical MD (top panel) and PIMD (bottom panel) simulations.
} \end{figure}

The identification of different layers of the water/air system was done using a scheme for the identification of truly interfacial molecules (ITIM) \cite{itim,gitim}. This algorithm uses a probe sphere to detect the molecules at the surface. The radius of the probe sphere was set as 0.2 nm, which has been proven to be a good value for water \cite{gitim}. A distance-based cluster search was also performed using cutoff value of 0.35 nm, which corresponds to the first minimum within the oxygen-oxygen radial distribution function. This search prevents the identifying of evaporated molecules as surface ones. Once the topmost layer is identified, the whole procedure is repeated, thus being capable of labeling each molecule belonging to the successive layers. We have used three layers in this work. Fig.~\ref{layers} shows the definition of the layers for subsequent analyses: L1 corresponds to the surface molecules, L2 considers both the surface and the immediate second molecular layer below the surface, whereas L3 includes surface molecules and molecules in two consecutive molecular layers. Finally, the whole system is considered as bulk water.
Once the surface molecules are identified for every frame of the simulation, the intrinsic mass density profile of water is obtained according to the equation
\begin{equation}
\rho(z)=\frac{1}{A}\left< \sum_i \delta(z-z_i+\xi(x_i,y_i))\right>,
\end{equation}
where $z_i$ is the position of the i-th particle and $\xi(x_i,y_i)$ contains the location of the water-air interface in the z-direction. Fig.~\ref{Fig2} shows the resulting intrinsic profile including the ones of the first three consecutive molecular layers. The first layer shows a Dirac delta peak at $z=0$, given that all molecules in this layer are at the surface. This figure shows rich features up to about 4 \AA\,, which would have been spread out in an averaged density profile. We note that the inclusion of NQEs does not alter the intrinsic density profile at the interface, which is in agreement with our previous work \cite{kessler}.

\begin{figure}
\begin{center}
\includegraphics[width=0.4875\textwidth]{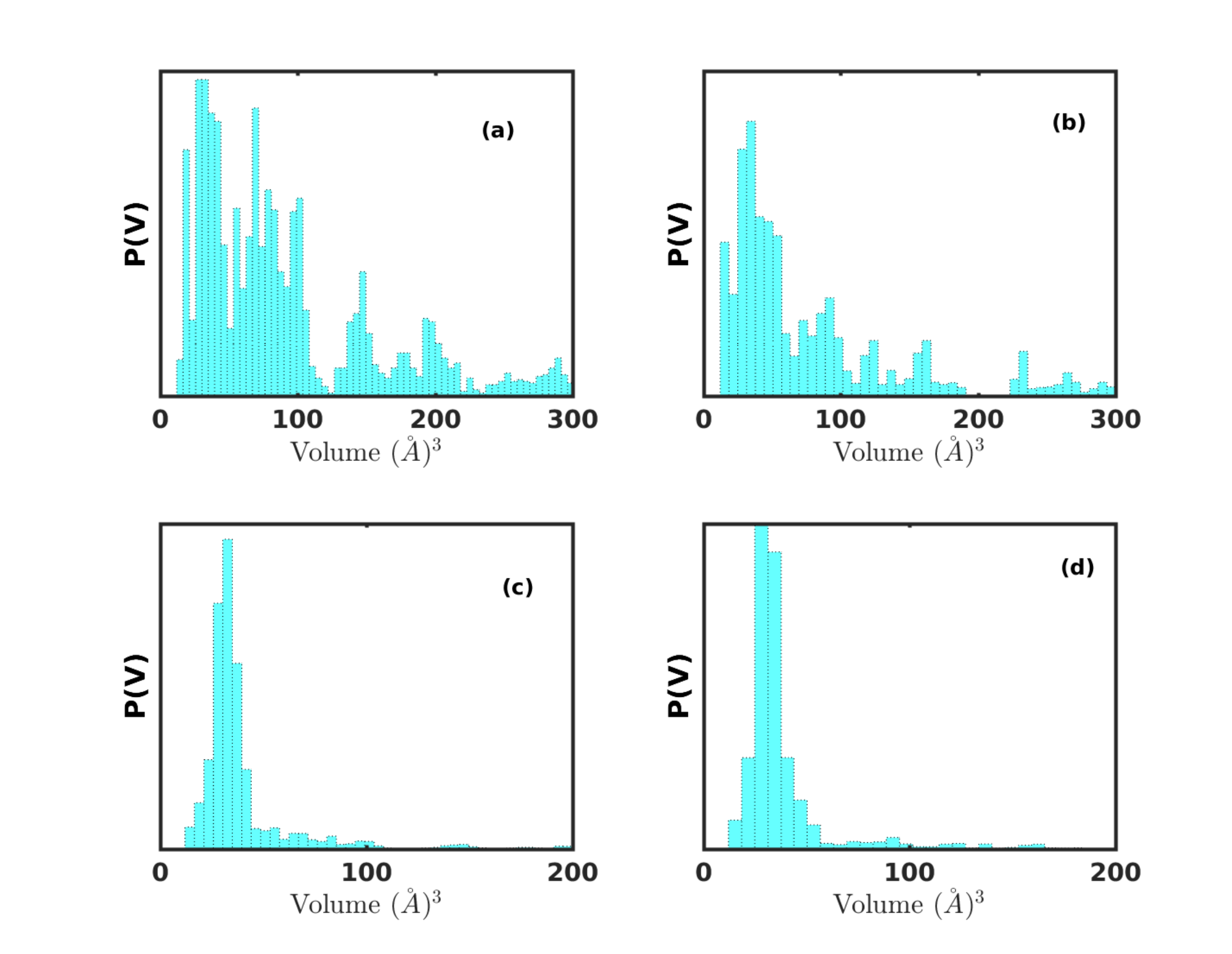}
\end{center}
\caption{\label{Fig3}
The Voronoi polyhedra distribution for water molecules present in (a) layer L1  obtained from PIMD, (b) layer L1 obtained from classical MD,
(c) bulk obtained from PIMD, and (d) bulk obtained from classical MD simulations.
} \end{figure}

We have analyzed the local structure at the water-air interface and bulk using Voronoi polyhedra analysis \cite{vornoi}. The Voronoi region of an atom is the region in space to which all the points are closer than to any other atom.  The distribution of Voronoi polyhedra volume for water molecules at layer L1 and for the bulk are shown in Fig.~\ref{Fig3} for both classical MD and PIMD simulated systems.
It is evident for the distributions that there is a significant increase in the volume of Voronoi polyhedra at the interface in comparison to the bulk, which in principle implies that water molecules are less
compactly packed at the water-air interface than in bulk. Furthermore, the distribution volume of Voronoi polyhedra of interfacial water molecules obtained from PIMD is also relatively wider than from classical MD simulations.

The reorientational dynamics of water molecules at the surface, i.e. layer L1, is analyzed using the orientational time correlation function, 
which is given as
\begin{equation} \label{Eq1}
C_{2}^{\mu }(t) = \left \langle P_{2}\left [ \mu(0) \cdot \mu(t)  \right ] \right \rangle,
\end{equation}
where $\mu(t)$ is the orientation of the molecular dipole vector at time $t$ and $P_{2}$ is the second-order Legendre polynomial.
\begin{figure}
\begin{center}
\includegraphics[width=0.4875\textwidth]{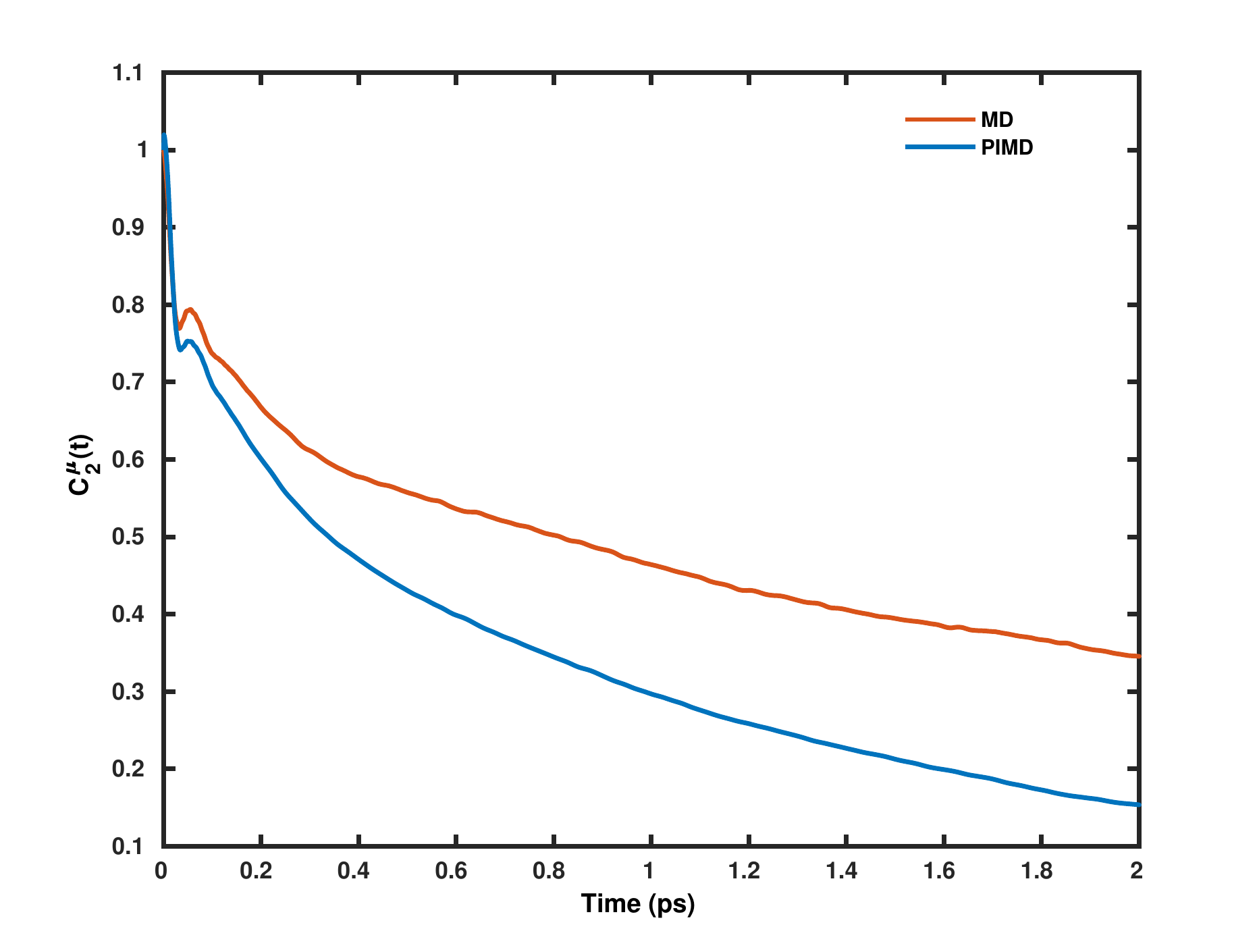}
\end{center}
\caption{\label{Fig4}
Orientational time correlation function of the molecular dipole vector of water molecules present in layer L1 as computed by MD and PIMD simulations.
} \end{figure}
The integrated time scale of the orientational time correlation function, which is shown in Fig.~\ref{Fig4}, gives the average time of the loss of correlation within the reorientational dynamics.
For the interfacial molecules in layer L1, the time scales of the orientational correlation function of the molecular dipole vectors are found to be 1.93~ps for the classical MD and 1.11~ps for the PIMD simulations. A recent study has shown that the water molecules on the water-air interface with free OH modes have a reorientational relaxation timescale of 0.81~ps \cite{yuki2}, which is in agreement with our PIMD results.

\subsection{Frequency fluctuation dynamics}
The time-dependent fluctuations in ground state frequencies of OH modes were calculated from PACMD trajectories using the wavelet method of  time-series analysis. The details for the calculation of OH frequencies have been described elsewhere \cite{ac5,ac6}. Here, we present a brief overview of the method. The
underlying principle of the wavelet method employed here is that a time-dependent function can be
expressed in terms of translations and dilations of a mother wavelet
\begin{equation} \label{1}
\psi_{a,  b}(t)=a^{-\frac{1}{2}}\psi\Bigg(\frac{t-b}{a}\Bigg),
\end{equation}
where the coefficients of the wavelet expansion are given by the wavelet transform of $f(t)$, i.e.
\begin{equation} \label{2}
L_{\psi}f(a,  b)=a^{-\frac{1}{2}}\int_{-\infty}^{+\infty}f(t)\bar{\psi}
\Bigg(\frac{t-b}{a}\Bigg)dt, 
\end{equation}
with $a$ and $b$ being real quantities with  $a$$>0$. 

We have employed the so-called
Morlet-Grossman form for the mother wavelet \cite{carmona}. The inverse of the scale factor $a$  is proportional
to the frequency
over a time window around $b$. The time-dependent function $f(t)$ is constructed as a complex function, with
its real and imaginary parts corresponding to instantaneous OH  bond distance and momentum of an OH mode
projected along the OH bond. This method has been applied to all the OH modes present in a given system.

\begin{figure}
\begin{center}
\includegraphics[width=0.4875\textwidth]{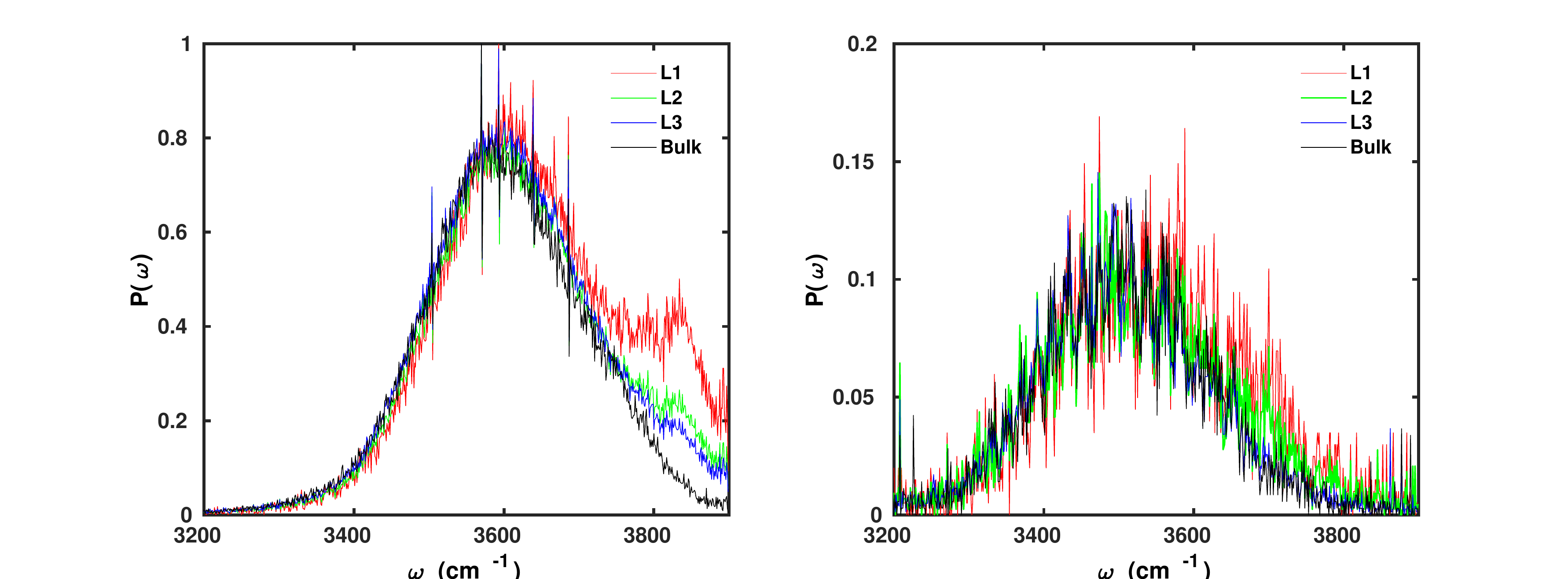}
\end{center}
\caption{\label{Fig5}
Vibrational frequency distributions of the O-H stretching modes in layer L1, L2, L3 and the bulk as obtained  from (a) classical MD and (b) PIMD simulations. 
} \end{figure} 
The distribution of OH frequencies as a function of position within the water-air interface has been analyzed for both systems and shown in Fig.~\ref{Fig5} (a) and (b), respectively.
We observe that the distribution of OH stretch frequencies corresponding to the layer L1 shows a distinct shoulder in the high-frequency regime of 3800-3900 cm$^{-1}$ from 
classical simulations. The shoulder is less prominent in the case of PIMD simulations, although an overall blueshift is still observable. The mean frequency of OH stretch modes
in layer L1 obtained from classical MD and quantum mechanical PIMD simulation are 3640  cm$^{-1}$ and 3518  cm$^{-1}$, respectively. Similarly, the frequency of OH stretch modes averaged over molecules in layer L2, comprising the two topmost layers of the water-air interface, as obtained from the classical and PIMD simulations are 3620  cm$^{-1}$ and 3502 cm$^{-1}$.
The distribution of OH stretch modes in the L3 layer is nearly indistinguishable with that of the bulk for both of the cases, with the average frequencies for L3  being 3612 and 3496  cm$^{-1}$, wheras the mean frequencies for all the OH modes are 3608 and 3493  cm$^{-1}$, respectively.
Purely from a static perspective, this implies that the water-air interface is characterized by the two topmost layers of water molecules.
Moreover, a blueshift of $\sim  $ 25-30  cm$^{-1}$ is observed in the mean frequency of OH modes present in layer L1  as compared to bulk water.
 A persistent redshift of 122  cm$^{-1}$ is seen for the interfacial, as well as bulk water molecules upon inclusion of NQEs.
 
\begin{figure}
\begin{center}
\includegraphics[width=0.4875\textwidth]{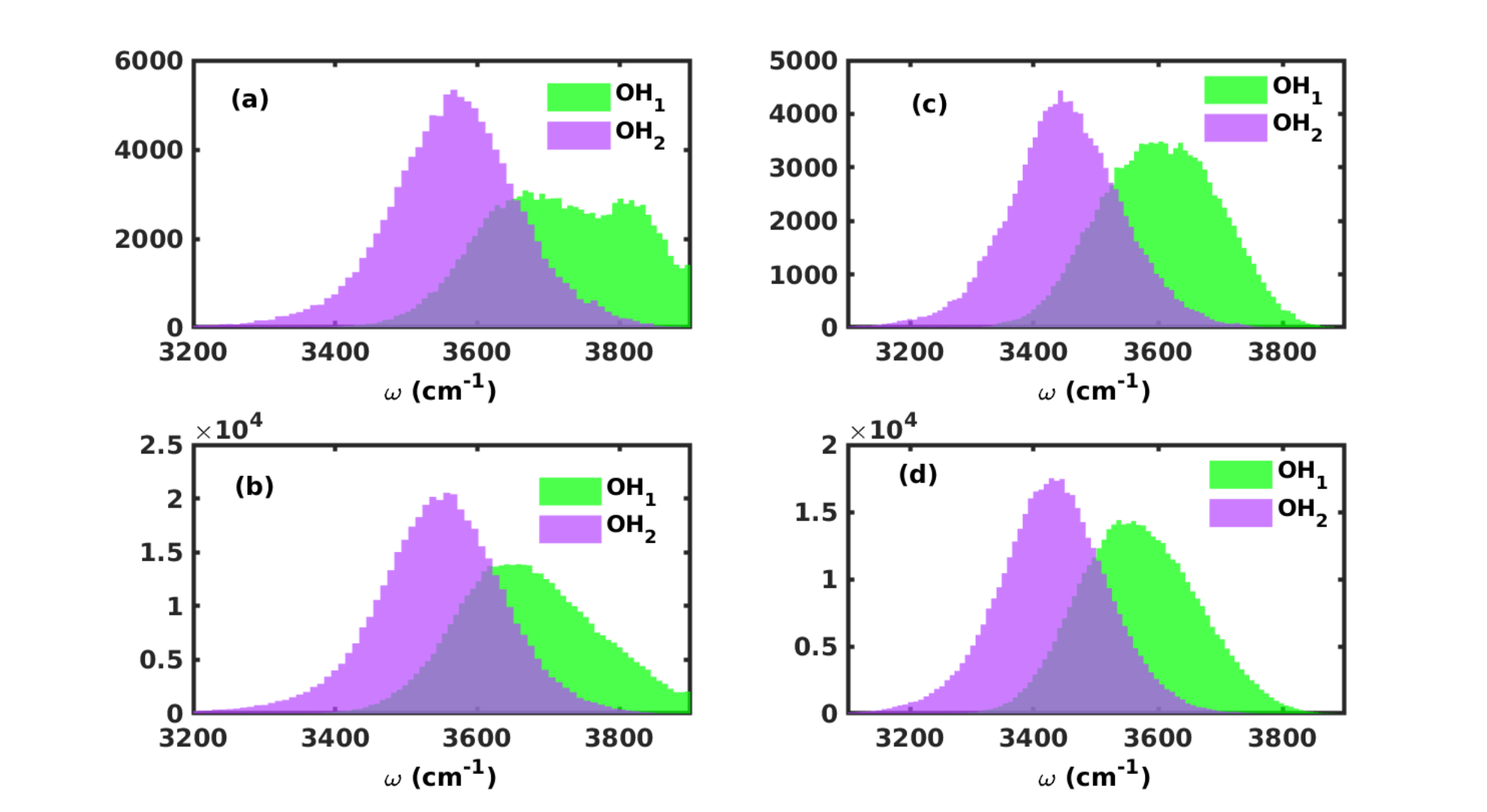}
\end{center}
\caption{\label{Fig6}
Vibrational frequency distributions of the stronger  ($OH_{1})$) and weaker ($OH_{2}$) intramolecular OH modes of water molecules  in (a) layer L1 and (b) bulk, as obtained by classical MD simulations, as well as (c) layer L1 and (d) bulk from quantum mechanical PIMD simulations.
} \end{figure}

\begin{figure}
\begin{center}
\includegraphics[width=0.4875\textwidth]{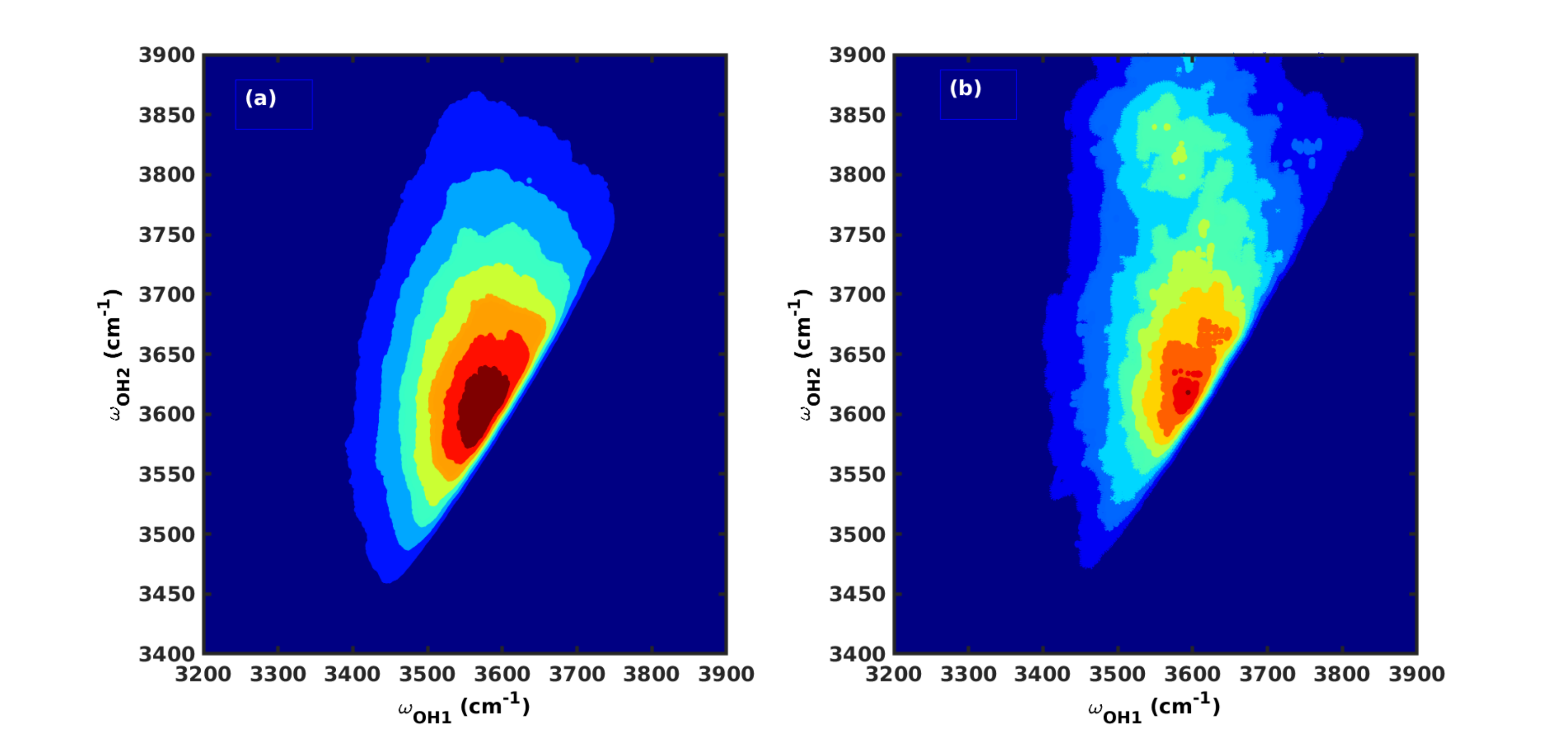}
\end{center}
\caption{\label{Fig7}
Joint probability distribution functions of the  stronger ($OH_{1})$) and weaker ($OH_{2}$) intramolecular OH modes of water molecules in (a) layer L1 and (b) in bulk, as obtained by classical MD simulations.
} \end{figure}

 Water molecules in liquid state are known to exhibit strong electronic asymmetry, which is a consequence of difference between the strength of first and second strongest hydrogen bond formed by a solvated water molecule \cite{Rustam1,Rustam2}. Since, the hydrogen bond strength is directly related to the vibrational frequencies of the OH modes in liquid water, we have analyzed the asymmetry between the intramolecular OH modes of water molecules in the bulk, as well as in layer L1 of the interface. The vibrational asymmetry is defined as the difference between the two vibrational frequency of the intramolecular OH modes of a given water molecule. We note that a vibrational asymmetry of 125 $cm^{-1}$ exists in bulk as obtained from our classical MD simulation. As the water molecules directly at the water-air interface are exposed to a more non-uniform solvent environment, the observed asymmetry gets amplified to 150 $cm^{-1}$. Interestingly, including NQEs by means of quantum mechanical PIMD simulations, the observed asymmetry between the intramolecular OH modes in bulk is with 140 $cm^{-1}$ slightly enhanced with respect to the corresponding classical simulations. Similarly, with the inclusion of NQEs, the interfacial water molecules experiences only a marginal increase in asymmetry of just 10 $cm^{-1}$ as compared to its classical counterpart with 150 $cm^{-1}$, as shown in Fig.~\ref{Fig6}. Moreover, the joint probability distribution functions of the stronger versus the weaker intramolecular OH mode for bulk and interfacial water, as obtained by the present classical MD simulations are also in Fig.~\ref{Fig7}. The observed asymmetric probability distribution further supports that the intramolecular OH mode asymmetry is more prominent at the water-air interface as compared to bulk.

\begin{figure}
\begin{center}
\includegraphics[width=0.4875\textwidth]{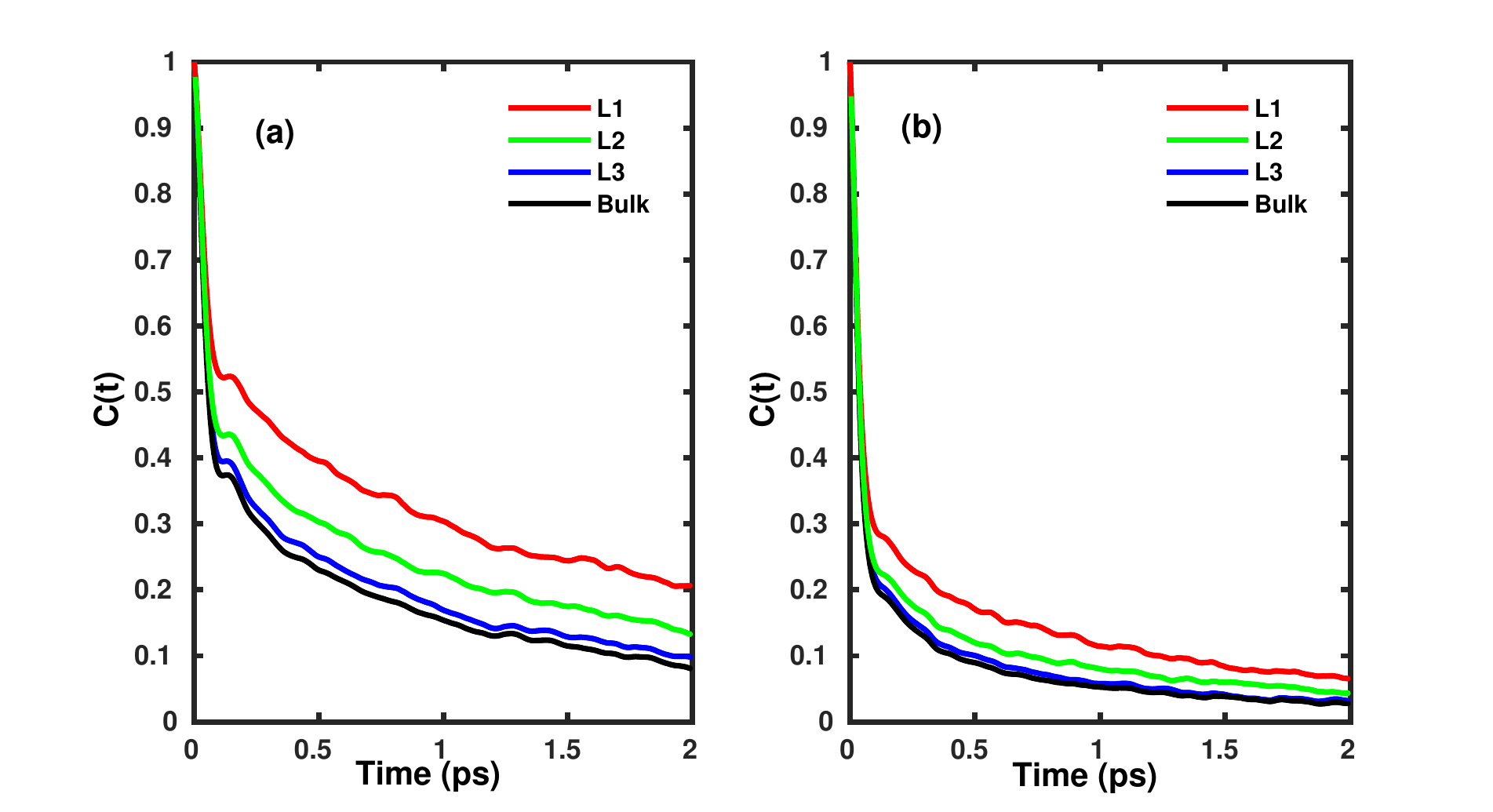}
\end{center}
\caption{\label{Fig8}
Time-dependent decay of the frequency-time correlation function of OH modes present in layer L1, L2, L3 and bulk, as obtained by (a) classical MD and (b) quantum mechanical PIMD simulations. 
} \end{figure}

Similarly, the dynamical evolution of vibrational frequencies of OH modes present in different layers of the water-air interface is determined using FTCFs. The time-dependent 
 decay of FTCFs for different layers and the bulk are shown in Fig.~\ref{Fig8} (a) and (b), respectively. The FTCFs for all cases show an initial fast decay which extends up to the first 100~fs, followed by a slow inertial decay which decays within a few picoseconds. The timescales of correlation decay were obtained from the raw data by using the following biexponential fitting function
\begin{equation}
 \mathit{f(t)}= a_{0}\exp\left(-\frac{t}{\tau _{0}}\right) + (1-a_{0})\exp\left(-\frac{t}{\tau _{1}}\right).
\end{equation}

\begin{figure}
\begin{center}
\includegraphics[width=0.4875\textwidth]{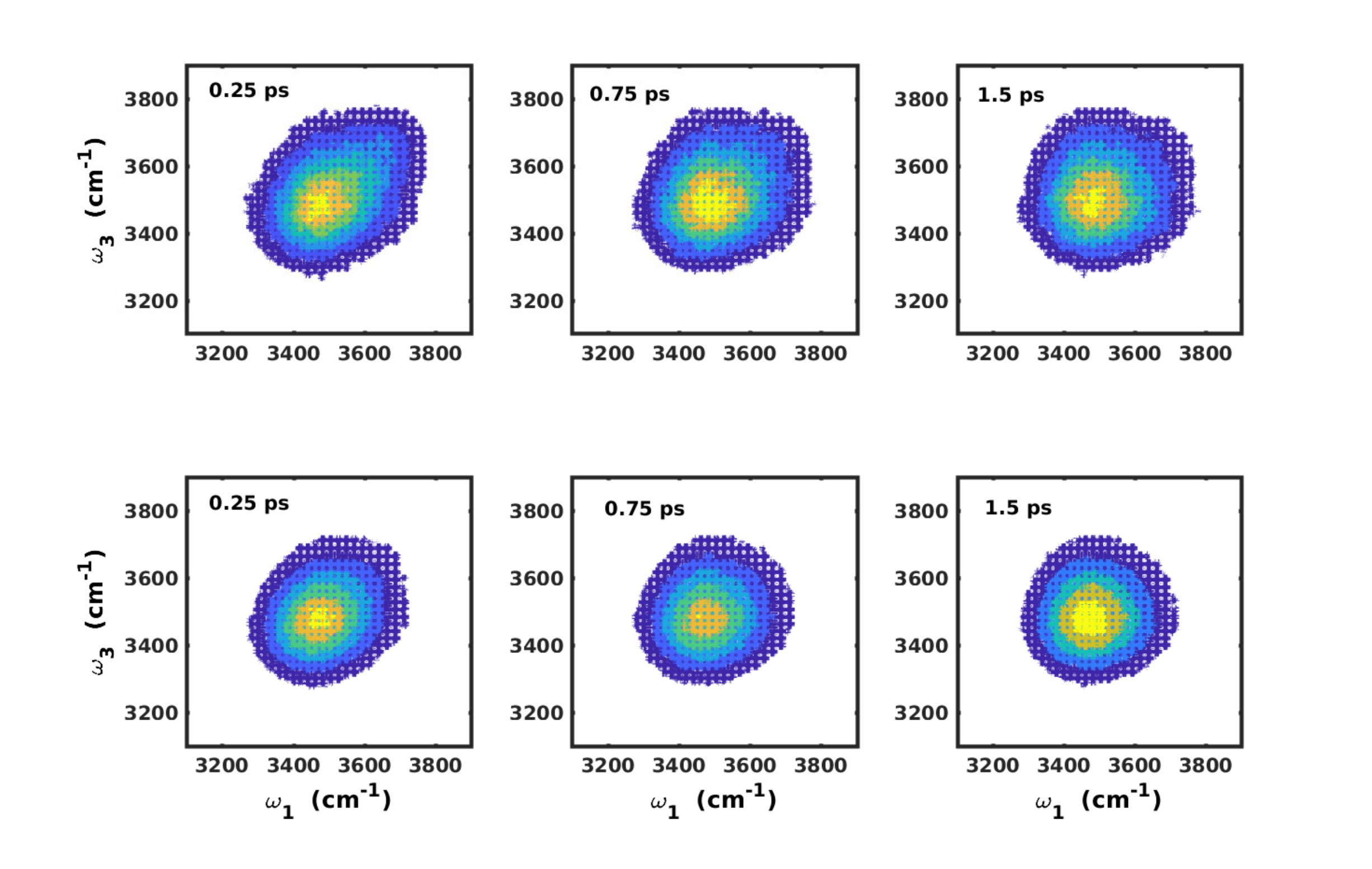}
\end{center}
\caption{\label{Fig9}
Joint probability distributions of finding the O-H stretching frequencies $\omega_{1}$ and $\omega_{3}$  at a time delay of $t_{2}$ for water molecules in layer L1 (top) and bulk  (bottom) obtained from PIMD simulations.
} \end{figure}

The timescale of decay of the long-tail component for layer L1, L2, L3 and bulk obtained from the classical simulations are 2.37, 1.98, 1.70 and 1.50~ps, respectively. 
In the case of PIMD, the timescale of decay of the long-tail component for layer L1, L2, L3, and bulk are 1.45, 1.31, 1.20 and 1.02~ps, respectively. It is evident that the dynamics of 
OH modes present in layer L1 with reference to bulk is slower by nearly 38 and 58 $\%$  for classical and PIMD simulations, respectively. Furthermore, the decay of layer L3 and bulk are found to be similar for both the initial fast and long-tail components in both simulations. It is indicative of the fact that from both a static and dynamical perspective, the water molecules beyond the top two layers are indistinguishable from bulk molecules. With reference to the NQEs, it is known that the vibrational dynamics of liquid water become accelerated by 30  $\%$  on inclusion of NQEs. However, for interfacial water molecules, these effects can lead to nearly 63 $\%$ acceleration. Along similar lines, we have also calculated the joint probability distribution of vibrational frequencies for OH modes of layer L1 and bulk  obtained from the 
PIMD simulation for different waiting times of 250, 750 and 1500~fs, as shown in Fig.~\ref{Fig9}. These distributions relax qualitatively at the timescale corresponding to the timescale of loss of frequency correlation.  It is evident from the plot that for bulk corresponding to the waiting time of 1500~fs frequency, the distribution becomes completely delocalized and spherical along both of the frequency axes. However, for the layer L1, the distribution still shows some asymmetric alignment along the diagonal, indicating slower vibrational spectral diffusion. 

\begin{figure}
\begin{center}
\includegraphics[width=0.4875\textwidth]{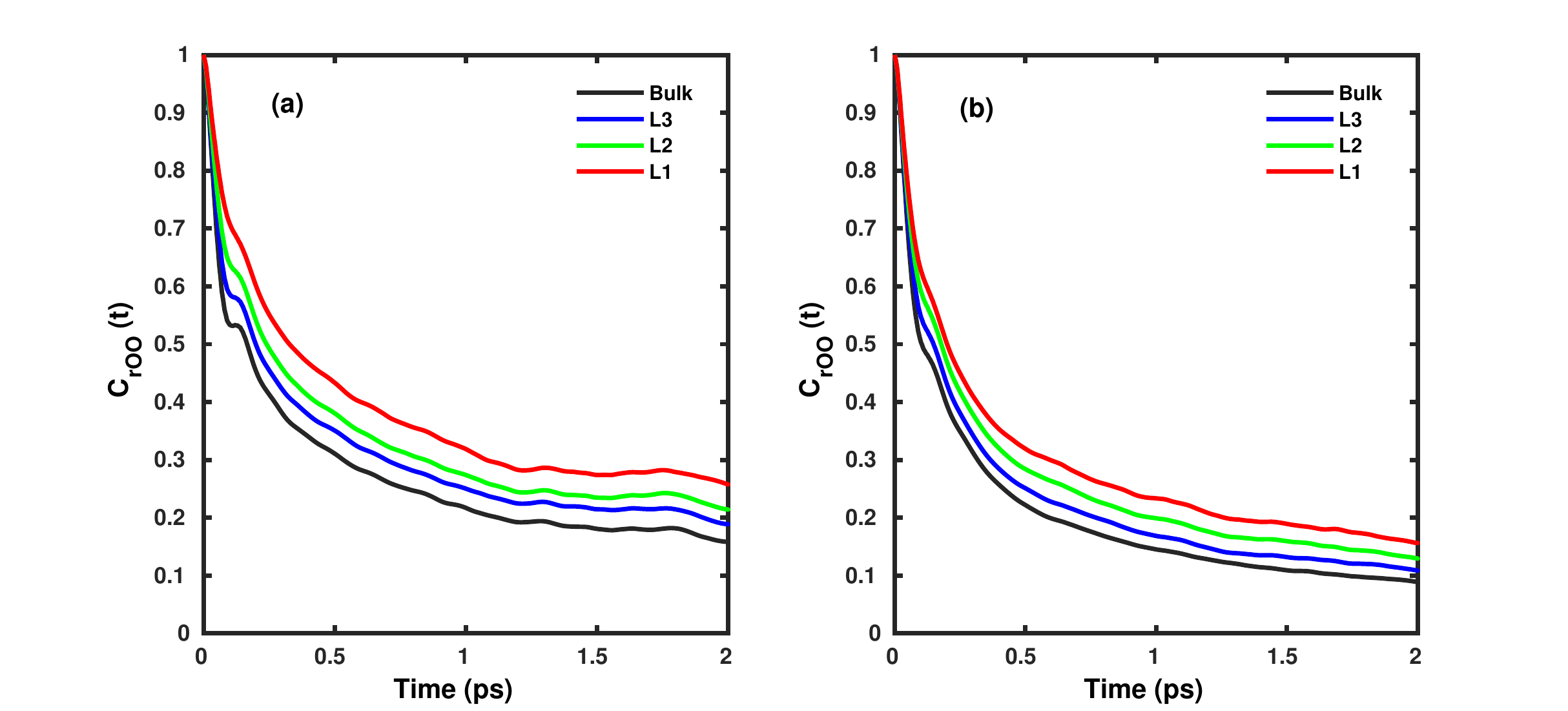}
\end{center}
\caption{\label{Fig10}
The time-dependent decay of the instantaneous fluctuations with respect to  $R_{O \cdots O}$ for OH modes present in layer L1, L2, L3 and bulk as determined by (a) classical MD and (b) PIMD simulations.
} \end{figure}

The instantaneous fluctuations in the vibrational frequencies of the OH modes in liquid water are strongly connected to the modulations in the local hydrogen bond structure. Accordingly,  we have calculated the time-dependent decay of fluctuations in $R_{O \cdots O }$ 
distance, i.e. $\left \langle \delta R_{O \cdots O}(0) \cdot \delta R_{O\cdots O}(t) \right \rangle$,  for the OH modes in the layers L1, L2, L3 and bulk for the classical and PIMD simulations, as shown in Fig.~\ref{Fig10}. The timescales are calculated by the least-squares fit to a biexponential function in analogy to Eq.~5. The longtime decay constant $ \tau_{1}$ is attributed to the timescale of loss of local structural correlation. The timescales of loss of structural correlation for the layers L1, L2, L3 and bulk obtained from classical simulations are 3.50, 2.70, 2.29 and 1.95~ps, respectively. Explicitly including NQEs leads to an overall accelerated dynamics, with resultant timescales for L1, L2, L3 and bulk being 2.40, 2.06, 1.90 and 1.64~ps, respectively. We observe that although the dynamics are slow compared to the decay of frequency correlation, the overall trends remain similar. Furthermore, the NQE contribution to the acceleration of vibrational dynamics of interfacial water molecules in layer L1 is 45 $ \%$. 

\begin{figure}
\begin{center}
\includegraphics[width=0.4875\textwidth]{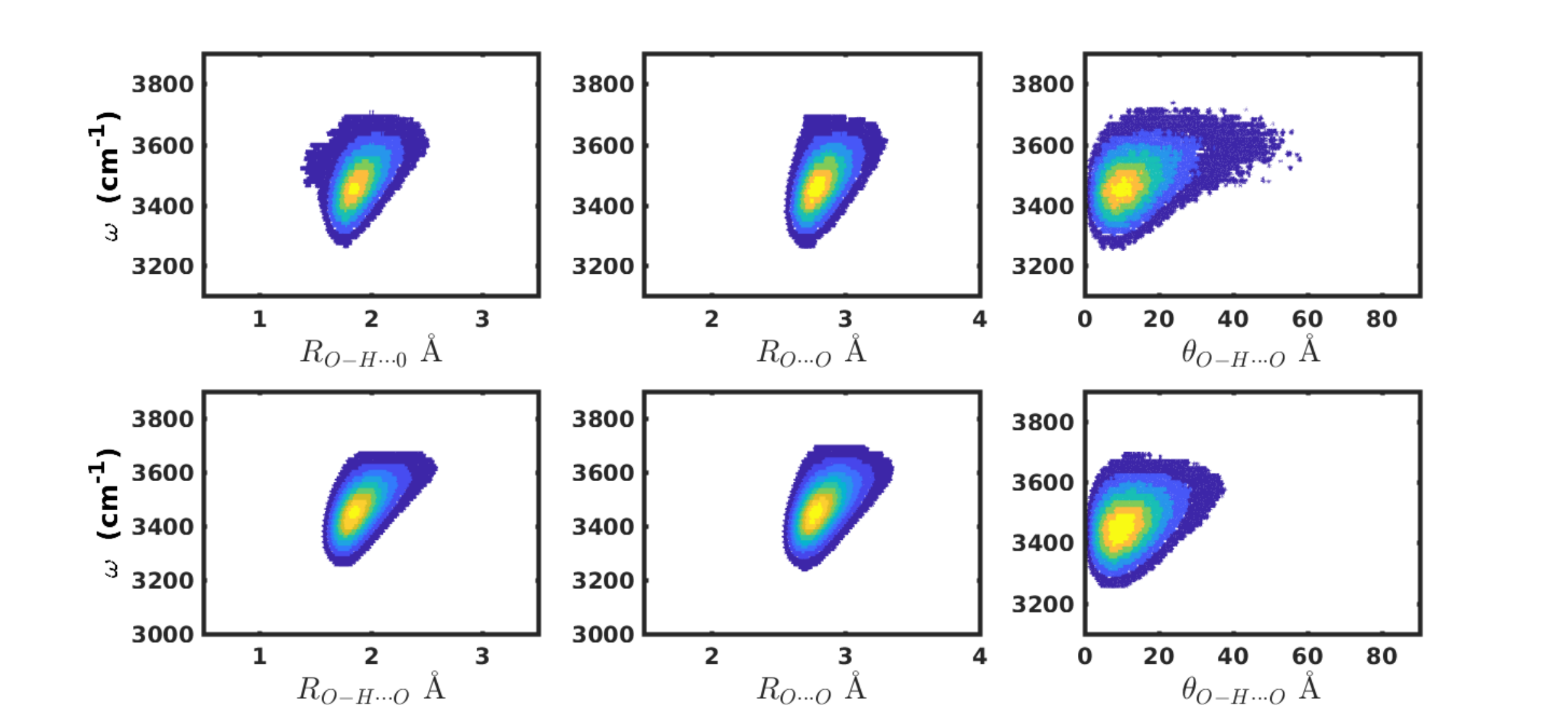}
\end{center}
\caption{\label{Fig11}
Frequency structure correlation between the instantaneous vibrational frequency of the O-H stretch mode as a function of $R_{H-O \cdots H}$, $R_{O \cdots O}$ and the angle $\theta_{O-H \cdots O}$ for layer L1 (top panel) and bulk (bottom panel) obtained from PIMD simulations.
} \end{figure}

While demonstrating that the temporal evolution of vibrational frequencies of OH modes is  correlated with the instantaneous fluctuations in the 
 $R_{H-O \cdots H}$ distance both in the bulk and at the interface, we also analyzed the spatial correlation between the modulations
in vibrational frequencies as a function of $R_{O \cdots H}$, $R_{O \cdots O}$ and $\theta_{O-H \cdots O}$ for water molecules in bulk and at 
interface, as shown in Fig.~\ref{Fig11}. Deviations of the OH modes at the water-air interface from the bulk behavior are easily seen in the region corresponding to higher vibrational frequencies. The increased  spread of $R_{O \cdots H}$ distance, $R_{O \cdots O}$ and $\theta_{O-H \cdots O}$
at the interface implies a less organized local hydrogen bond network that is  less correlated with the 
fluctuations of the OH stretch modes in comparison to the bulk.

\begin{figure}
\begin{center}
\includegraphics[width=0.4875\textwidth]{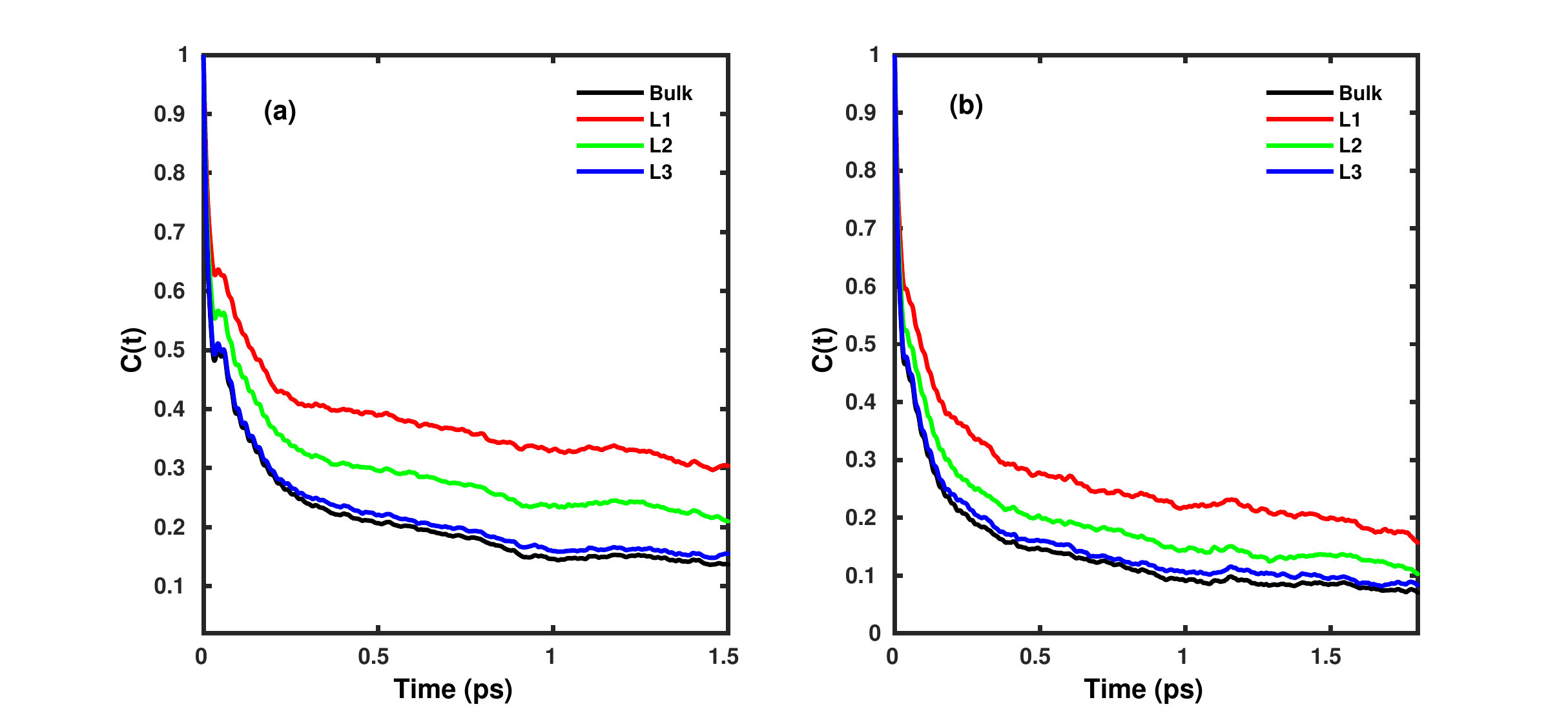}
\end{center}
\caption{\label{Fig12}
The time-dependent decay of the hydrogen bond number correlation function of water molecules in layer L1, L2, L3 and bulk, as obtained from (a) classical MD and (b) PIMD simulations.
} \end{figure}

The frequency correlation decay of water molecules is strongly influenced by the reorganization of the local hydrogen bond network. In order to correlate the 
vibrational correlation loss with the hydrogen bond network, we have calculated the hydrogen bond number correlation function, which shown in Fig.~\ref{Fig12}, and defined as
\begin{equation} 
C(t) = \frac{\left \langle \delta n(t) \cdot \delta n(0) \right \rangle}{\left \langle \delta n^2 \right \rangle},
\end{equation}
where $\delta n(t) = n(t) -\left \langle n \right \rangle$ and $n(t)$ is the number of hydrogen bonds that a water molecule forms at time instant $t$.
The timescales of decay of the 
layer-specific correlation functions were determined using a biexponential least-squares fitting function of Eq.~5. The longer timescale $ \tau_{1}$ is attributed to the loss of correlation in the average hydrogen bond number. The timescales for the layer L1, L2, L3 and the bulk from our classical MD simulations were found to be 3.40, 2.73, 2.40 and 1.90~ps, respectively. The inclusion of NQEs leads to an overall accelerated dynamics, with timescales for L1, L2, L3 and bulk water of 
2.12, 1.66, 1.45 and 1.27~ps, respectively.
Furthermore, here we also observe an acceleration by a factor of nearly 60 $\%$ with respect to the vibrational dynamics at the interface.

\begin{figure}
\begin{center}
\includegraphics[width=0.4875\textwidth]{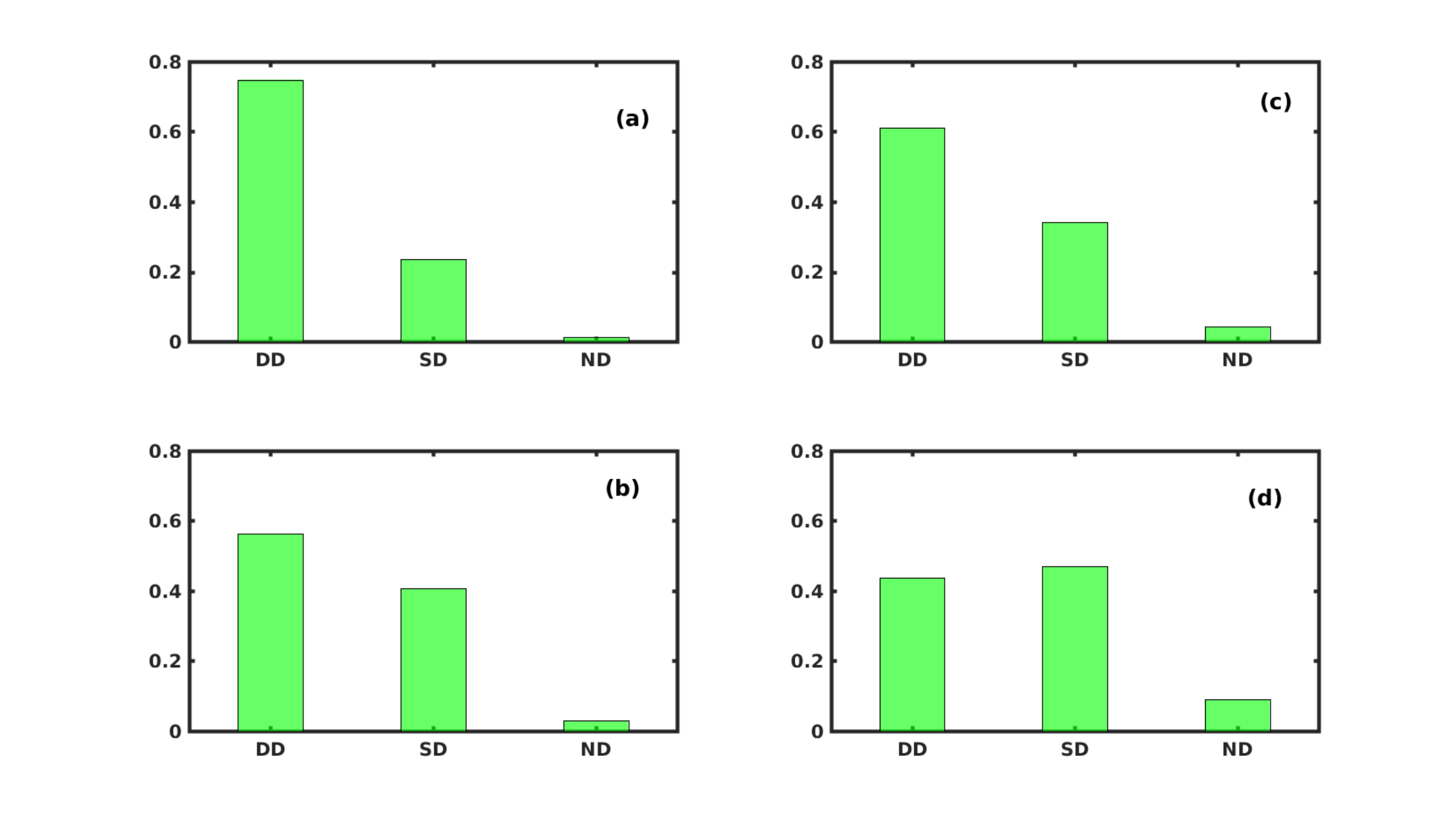}
\end{center}
\caption{\label{Fig13}
The population distribution of water molecules in different hydrogen bonding states in (a) layer L1 and (b) bulk obtained from classical MD simulation, as well as (c) layer L1 and (d) bulk obtained from PIMD simulations.
} \end{figure}

While the vibrational correlation loss is strongly correlated with the reorganization of the hydrogen bond network, the increased timescales of vibrational spectral diffusion at the water-air interface in comparison to bulk indicates significant differences in time-averaged hydrogen bonding patterns at the interface. We have calculated the population of water molecules in different 
hydrogen bonding states, namely the $(i)$ DD state (with both OH modes hydrogen-bonded), $(ii)$ SD state (with one of the OH modes hydrogen-bonded and the other dangling), and $(iii)$
ND state (with free OH modes). The population distribution of DD, SD and ND states in the bulk from the classical simulation was found to be 0.74, 0.23 and 0.01, respectively. With NQEs, the population distribution for the bulk is modified to 0.61, 0.34 and 0.04, respectively. We note that explicitly accounting for NQEs leads to a similar population distribution, but an evidently less structured hydrogen bond network, as seen through the increase in the population of the SD state.  The population distribution of DD, SD and ND states at the water-air interface obtained from our classical MD simulations is 0.56, 0.40 and 0.03, respectively. Upon considering NQEs, the population of the three states becomes 0.43, 0.47 and 0.09, respectively. Thus, we note that the increased population of the SD state is responsible for the overall slowed down
decay of frequency correlation at the interface. With reference to the inclusion of NQEs, we find that the majority of water molecules are predicted to be in the SD state. The population distribution for the bulk and the interface are both shown in Fig.~\ref{Fig13}.

\begin{figure}
\begin{center}
\includegraphics[width=0.4875\textwidth]{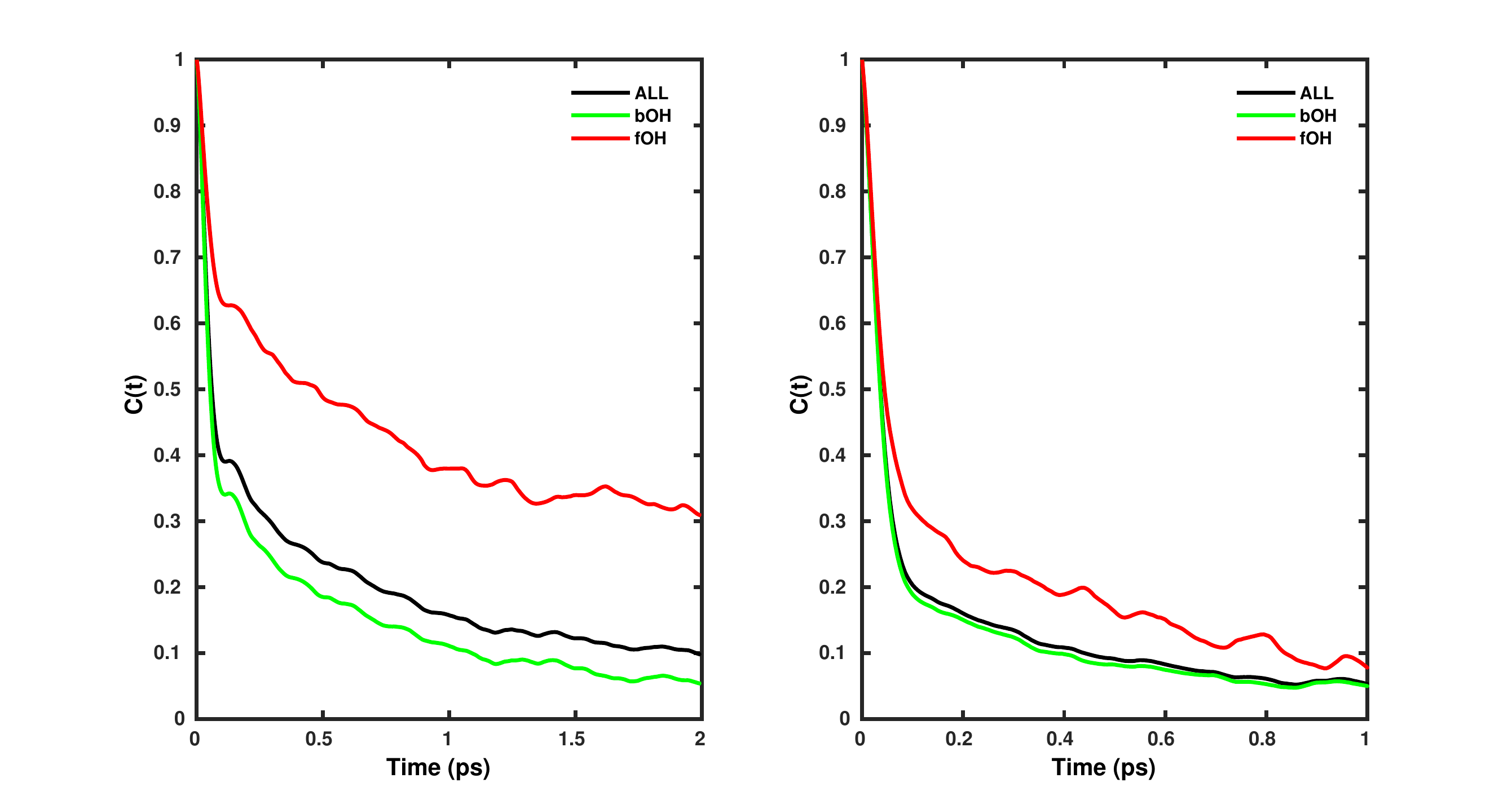}
\end{center}
\caption{\label{Fig14}
Time-dependent decay of frequency correlation of free OH (fOH) and bonded OH (bOH) modes, as obtained from (a) classical MD and (b) PIMD simulations. The time-averaged decay of all OH modes (in black) is for comparison.
} \end{figure}

As the population distribution is increased at the interface, we calculated the configuration-wise frequency correlation decay for all OH modes present in the system. The OH modes were classified into two classes: free OH (fOH), i.e. SD and ND configurations, and bonded OH (bOH), which correspondins to the DD state. 
The timescale of decay of FTCFs were obtained using the biexponential fitting function, as shown in Eq.~5. The timescale of decay of fOH and bOH for the PIMD simulated system were 1.20 and 0.74~ps, respectively.
Similarly, for the classical case, the timescales obtained for fOH and bOH modes were found to be 2.78 and 1.00~ps, respectively. We note a striking similarity in the timescales of decay of fOH modes and the timescales of decay of the FTCF for layer L1, which evidently implies that the slowdown of vibrational relaxation is predominantly due to the increased population of fOH modes on the surface that characteristically have a slower decay of frequency correlation. All results for the fOH and bOH modes are shown in Fig.~\ref{Fig14}. The correlation decay of all modes is also shown for comparison.

\begin{figure}
\begin{center}
\includegraphics[width=0.50\textwidth]{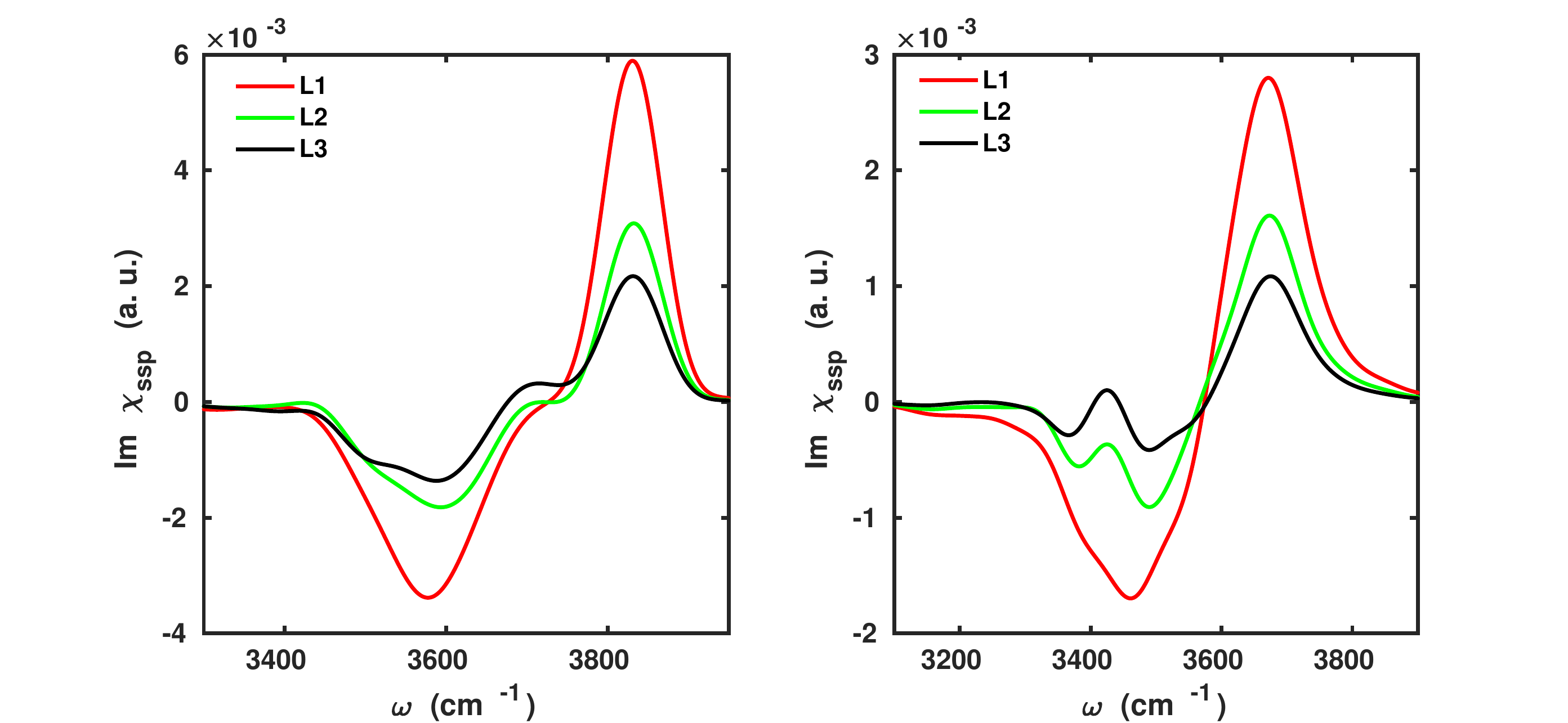}
\end{center}
\caption{\label{Fig15}
vSFG of OH modes present in layer L1, L2 and L3 determined from (a) PIMD and (b) classical MD simulations.
} \end{figure}

The vSFG spectrum of the water-air interface is calculated for the layers L1, L2 and L3 for both of the systems, as shown in Fig.~\ref{Fig15}. The vSFG was calculated using   the surface-specific velocity-velocity correlation function approach \cite{yuki,tdk_cc,tdk_sr2}, i.e.
\begin{equation*} \label{Eq4}
\chi^{2}_{abc}(\omega) = \left\{\begin{matrix}
\frac{Q(\omega) \mu^{'}_{str}\alpha^{'}_{str}  }{i\omega^{2} } \\ \int_{0}^{\infty}dte^{i\omega t}\left \langle \sum_{i,j}\dot{r}^{OH}_{c,j}(0)\frac{\overrightarrow{\dot{r}}^{OH}_{j}(t).\overrightarrow{r}^{OH}_{j}(t)}{\left | \overrightarrow{r}^{OH}_{j}(t) \right |}  \right \rangle & a=b \\ 
0 & a \neq b.
\end{matrix}\right.
\end{equation*}
where $\chi_{2}^{abc}$ is the resonant component of susceptibility, while $r^{OH}_{j}$ and $\dot{r}^{OH}_{j}$ refer to the intramolecular distance and velocity of a given OH mode, respectively.
With respect to the contributions of the individual layers to the vSFG intensity, it is seen that as we move below the water-air surface, the contributions become negligible,  which is in agreement with the decay trends of FTCF for the three layers. Furthermore, the explicit inclusion of NQEs leads to an apparent redshift in the peak positions and an overall broadening of the peaks corresponding to the free and bonded OH modes in comparison to the classical MD simulations. 

\section{Summary}
To summarize, the impact of NQEs on the vibrational dynamics of water molecules directly at the liquid-vapor interface is investigated in detail by means of PIMD simulations. We find that the explicit inclusion of NQEs leads to a reduction of frequency correlation timescales by as much as 60~$\%$. While an overall reduction in timescales is easily discerned, a significantly faster decay in the initial diffusive regime is also noticeable. Furthermore, it is also found that the static vibrational distribution, as well as the temporal decay of frequency correlation of water molecules beyond the two topmost layers of molecules is essentially identical to that of bulk water. A redshift of 120 cm$^{-1}$ is noticed for both the interfacial and bulk water molecules upon the inclusion of NQEs. The water molecules on the surface show a blueshift due to the increased population of SD configurations at the interface. The timescales of decay of the hydrogen bond number correlation and local structure correlation also follow the observed decay pattern of frequency correlation loss. The computed vSFG spectra leads to similar inferences.

\begin{acknowledgments}
The authors would like to thank the Paderborn Center for Parallel Computing (PC$^2$) for the generous allocation of computing time on FPGA-based supercomputer ``Noctua'', as well as the Gauss Center for Supercomputing (GCS) for providing computing time through the John von Neumann Institute for Computing (NIC) on the GCS share of the supercomputer JUWELS at the J\"ulich Supercomputing Centre (JSC). This project has received funding from the European Research Council (ERC) under the European Union's Horizon 2020 research and innovation programme (grant agreement No 716142).
\end{acknowledgments}


\begin{thebibliography}{10}

\bibitem{kauzmann}
D. S. Eisenberg and W. Kauzmann
\newblock {\em The Structure and Properties of Water}
\newblock  (Oxford University Press,  1999).

\bibitem{ball}
P.~Ball,
\newblock {\em Life's Matrix: A Biography of Water}
\newblock (Univ of California Press 2001).

\bibitem{ball2008}
P. Ball,
\newblock {\em Chem. Rev.}  {\bf 108}, 74 (2008).



\bibitem{richmond3}
L. F. Scatena, M. G. Brown and G. L. Richmond, 
\newblock {\em Science} {\bf 292}, 5518 (2001).

\bibitem{richmond2}
G. L. Richmond,
\newblock {\em Annu. Rev. Phys. Chem.} {\bf 52}, 357 (2001).

\bibitem{richmond}
G. L. Richmond,
\newblock {\em  Chem. Rev.} {\bf 102}, 2693 (2002).

\bibitem{skinner}
H. J. Bakker and J. L. Skinner,
\newblock {\em  Chem. Rev.} {\bf 110}, 1498 (2010).

\bibitem{tdk_1}
T. D. K\"uhne, T. A. Pascal, E. Kaxiras and Y. Jung,
\newblock {\em  J. Phys. Chem. Lett.} {\bf 2}, 105 (2011).

\bibitem{tdk_4}
K. Karhan, R. Z. Khaliullin and T. D. K\"uhne,
\newblock {\em  J. Chem. Phys.} {\bf 141}, 22D528 (2014)

\bibitem{tdk_3}
Y. Nagata, T. Ohto, M. Bonn and T. D. K\"uhne,
\newblock {\em  J. Chem. Phys.} {\bf 144}, 204705 (2016).

\bibitem{skinner2}
J. L. Skinner, P. A. Pieniazek and S. M. Gruenbaum,
\newblock {\em  Chem. Rev.} {\bf 45}, 93 (2012).

\bibitem{tdk_sr}
D. Ojha, K. Karhan and T. D. K\"uhne, 
\newblock {\em Sci. Rep. } {\bf 8}, 16888 (2018).

\bibitem{shen1}
Q. Du, R. Superfine, E. Freysz and Y. R. Shen,
\newblock {\em  Phys. Rev. Lett.} {\bf 70}, 2313 (1993).

\bibitem{shen3}
V. Ostroverkhov, G. A. Waychunas and Y. R. Shen,
\newblock {\em  Phys. Rev. Lett.} {\bf 94}, 046102 (2005).

\bibitem{shen2}
Y. R. Shen and  V. Ostroverkhov,
\newblock {\em  Chem. Rev.} {\bf 106}, 1140 (2006).

\bibitem{mbonn}
J. Bredenbeck, A. Ghosh, H-K Nienhuys and M Bonn,
\newblock {\em  Acc. Chem. Res.} {\bf 42}, 1332 (2009).

\bibitem{allen}
D. Verreault, W. Hua and H. C. Allen,
\newblock {\em  J. Phys. Chem. Lett.} {\bf 20}, 3012 (2012).

\bibitem{tahara}
S. Nihonyanagi, S. Yamaguchi, T. Tahara, 
\newblock {\em  Chem. Rev.} {\bf 16}, 10665 (2017).

\bibitem{morita1}
A. Morita and J. T. Hynes,
\newblock {\em Chem. Phys.} {\bf 258}, 371  (2000).

\bibitem{morita2}
A. Morita and J. T. Hynes,
\newblock {\em J. Phys. Chem. B} {\bf 106}, 673 (2002).

\bibitem{paesani1}
G. R. Medders and F.  Paesani, 
\newblock {\em J. Amer. Chem. Soc.} {\bf 138}, 3912--3919 (2016).

\bibitem{richmond4}
E. A. Raymond, T. L. Tarbuck, G. L. Richmond,
\newblock {\em J. Phys. Chem. B} {\bf 106}, 2817 (2002).

\bibitem{richmond5}
E. A. Raymond, T. L. Tarbuck, G. L. Richmond,
\newblock {\em J. Phys. Chem. B} {\bf 107}, 546 (2003).

\bibitem{skinner3}
P. A. Pieniazek, C. J. Tainter and J. L. Skinner,
\newblock {\em J. Amer. Chem. Soc.} {\bf 133}, 10360 (2011).

\bibitem{skinner4}
B. M. Auer and J. L. Skinner,
\newblock {\em J. Chem. Phys.} {\bf 129}, 214705  (2008).

\bibitem{skinner5}
B. M. Auer and J. L. Skinner,
\newblock {\em J. Phys. Chem. B} {\bf 113}, 4125 (2009).

\bibitem{skinner6}
P. A. Pieniazek, C. J. Tainter and J. L. Skinner,
\newblock {\em J. Chem. Phys.} {\bf 135}, 044701  (2011).

\bibitem{skinner7}
Y. Li, M. Gruenbaum and J. L. Skinner,
\newblock {\em  Proc. Nat. Acad. Sci.} {\bf 110}, 1992 (2013).

\bibitem{yuki}
T. Ohta, K. Usui, T. Hasegawa, M. Bonn and Y. Nagata,
\newblock {\em J. Chem. Phys.} {\bf 143}, 124702  (2015).

\bibitem{tdk_cc}
D. Ojha, N. K. Kaliannan and T. D. K\"uhne, 
\newblock {\em Commun. Chem.} {\bf 2}, 116 (2019).

\bibitem{tdk_sr2}
D. Ojha and T. D. K\"uhne, 
\newblock {\em Sci. Rep.} {\bf 11}, 2456 (2021).

\bibitem{tdk_wp}
D. Ojha and T. D. K\"uhne, 
\newblock {\em Molecules} {\bf 25}, 3939 (2020).

\bibitem{kessler}
J. Kessler, H. Elgabarty, T. Spura, K. Karhan, P. Partovi-Azar, A. A. Hassanali and T. D. K\"uhne, 
\newblock {\em J. Phys. Chem. B} {\bf 119}, 10079 (2015).

\bibitem{naveen}
N. K. Kaliannan, A. Henao, H. Wiebeler, F. Zysk, T. Ohto, Y. Nagata and T. D. K\"uhne, 
\newblock {\em Mol. Phys.} {\bf 118}, 1620358 (2020).

\bibitem{sprik}
M.  Sulpizi, M. Salanne, M. Sprik and M. P.  Gaigeot, 
\newblock {\em J. Phys. Chem. Lett.} {\bf 4}, 83  (2013).

\bibitem{simone}
S. Pezzoti, R. Galimberti and M. P. Gaigeot,
\newblock {\em J. Phys. Chem. Lett.} {\bf 8}, 3133  (2017).

\bibitem{ceriotti}
M. Ceriotti,  W. Fang, P. G. Kusalik, R. H. Mckenzie, A. Michaelides,
M. A. Morales and T. E. Markland,
\newblock {\em  Chem. Rev.} {\bf 116}, 7529 (2016).


\bibitem{hibbs}
R. P. Feynman and A. R. Hibbs,
\newblock {\em Quantum Mechanics and Path Integrals}, {\em McGraw-Hill, New York} (1965).

\bibitem{PIMD}
M. Parrinello and A. Rahman,
\newblock {\em J. Chem. Phys.} {\bf 80}, 860 (1984).

\bibitem{paesani}
F. Paesani,  S. Iuchi and G. A. Voth,
\newblock {\em J. Chem. Phys.} {\bf 127}, 074506  (2006).

\bibitem{haberson_3}
S. Habershon, T. E. Markland and D. E. Manolopoulos,
\newblock {\em  J. Chem. Phys.} {\bf 131}, 24501 (2009).

\bibitem{tdk_5}
T. Spura, C. John, S. Habershon and T. D. K{\"u}hne,
\newblock {\em Mol. Phys.} {\bf 113}, 808 (2015).

\bibitem{laage}
D. M. Wilkins, D. E. Manolopoulos, S. Pipolo, D. Laage and J. T. Hynes,
\newblock {\em J. Phys. Chem. Lett.} {\bf 8}, 2602 (2017).

\bibitem{tdk_6}
D. Ojha, A. Henao and T. D. K\"uhne, 
\newblock {\em J. Chem. Phys.} {\bf 148}, 102328 (2018).

\bibitem{hone}
T. D. Hone, P. J. Rossky and G. A. Voth,
\newblock {\em  J. Chem. Phys.} {\bf 124}, 154103 (2006).

\bibitem{MTS}
M. E. Tuckerman, B. J. Berne and G. J. Martyna,
\newblock {\em J. Chem. Phys.} {\bf 97}, 1990 (1992).

\bibitem{itim}
L. B. P\'artay, G. Hantal, P. jedlovszky, \'A. Vincze and G. Horvai,
\newblock {\em  J. Comp. Chem.} {\bf 29}, 945 (2008).

\bibitem{gitim}
M. Sega, S. S. Kantorovich, P. Jedlovszky and M. Jorge,
\newblock {\em  J. Chem. Phys.} {\bf 138}, 044110 (2013).

\bibitem{vmd}
W. Humphrey, A. Dalke and K. Schulten,
\newblock {\em  J. Molec. Graphics} {\bf 14}, 33-38 (1996).

\bibitem{vornoi}
M. Tanemura, T. Ogawa and N. Ogita,
\newblock {\em  J. Comp. Phys.} {\bf 51}, 191 (1983).

\bibitem{yuki2}
J. Jeon, C. S.  Hsieh, Y. Nagata, M. Bonn and M. Cho,
\newblock {\em  J. Chem. Phys.} {\bf 147}, 044707 (2017).

\bibitem{ac5}
D. Ojha and A. Chandra,
\newblock {\em  J.  Comput. Chem.} {\bf 129},  (2019).

\bibitem{ac6}
D. Ojha and A. Chandra,
\newblock {\em  J.  Phys. Chem. B} {\bf 123}, 3325 (2019).

\bibitem{Rustam1}
T. D. K\"uhne and R. Z. Khaliullin,
\newblock {\em  Nature Commun.} {\bf 4}, 1450 (2013).

\bibitem{Rustam2}
T. D. K\"uhne and R. Z. Khaliullin,
\newblock {\em  J. Am. Chem. Soc.} {\bf 136}, 3395 (2014).


\bibitem{carmona}
R. Carmona, W. Hwang and  B. Torresani, \textit{Practical Time-frequency
Analysis: Gabor and Wavelet Transforms with an Implementation in S},
Academic Press, 1998.

\end{thebibliography}


\end{document}